\documentstyle[twocolumn,aps,prl]{revtex}

\begin{document}
\draft
\title{ Decimation Studies of Bloch Electrons in a Magnetic Field:
Higher Order Limit Cycles Underlying the Phase Diagram }
\author{ Jukka A. Ketoja\cite{address}, Indubala I. Satija\cite{email},
and Juan Carlos Chaves}
\address{
 Department of Physics and\\
 Institute of Computational Sciences and Informatics,\\
 George Mason University,\\
 Fairfax, VA 22030}
\date{\today}
\maketitle
\begin{abstract}
A decimation method is applied to the tight binding model
describing the two dimensional electron gas with next nearest
neighbor interaction
in the presence of an inverse golden mean magnetic flux.
The critical phase with fractal
spectrum and wave function exists in a finite window in two-dimensional
parameter space. There are special
points on the boundaries as well as inside the critical phase
where the renormalization flow exhibits
higher order limit cycles.
Our numerical results suggest
that most of the critical phase is characterized by a strange
attractor of the renormalization equations.
\end{abstract}

\pacs{75.30.Kz, 64.60.Ak, 64.60.Fr}
\narrowtext
\section{Introduction}

The two-dimensional electron gas with irrational magnetic flux is a
well-known paradigm in the study of systems with two competing periodicities.
The magnetic field results in reducing the problem to a one-dimensional
tight binding model (TBM) \cite{Sok} known as the Harper equation.\cite{Harper}
The Harper equation exhibits both extended (E) and localized (L) states.
At the onset of transition corresponding to a periodic potential
with square symmetry, the states are critical (C) with fractal spectra
and wave functions. The scaling properties of the devil staircase spectra
and the wave functions have been studied extensively using various
renormalization group (RG) methods.\cite{Ostlund},\cite{rg},\cite{KS}

Recently, it was pointed out that the Harper
equation also describes the isotropic XY quantum spin model in a
modulating magnetic field of periodicity incommensurate with the
periodicity of the lattice. \cite{IIS} It was shown that
the presence of anisotropy in spin space fattened the critical point
of the Harper equation resulting in a phase diagram where E, L, and C
phases all
existed in a finite measure parameter interval. The existence of a fat C phase
provided a new
scenario for the breakdown of analyticity in incommensurate systems.
Furthermore, based on numerical results obtained using
a new decimation method \cite{KS},\cite{Ketoja}
it was argued that the fat C phase was described by four distinct
universality classes characterized by limit cycles of the RG flow.

The fat C phase was also reported recently in
the TBM describing
Bloch electrons moving on a tight binding square lattice where the coupling
to next nearest neighbor (NNN) sites exceeded a certain threshold value
compared to the nearest neighbor (NN) coupling.\cite{Thou}
Using various analytical and numerical tools to study the scaling
properties of the fractal eigenspectrum, the universality
classes of the C phase were investigated. The E-C and C-L transition
lines were conclusively shown to define a new universality class,
different from the E-L transition line (which belonged to the Harper
class). However, the analysis in the interior of the C phase
was rather inconclusive.

In this paper, we apply decimation methods to the TBM describing Bloch
electrons in a field with
a NNN interaction. Unlike the previous study\cite{Thou} which
investigated the scaling properties of the fractal \underline{eigenvalues},
we study the scaling properties of the fractal \underline{eigenstates}.
Our study, in agreement with the previous work\cite{Thou},
shows that the E-C transition line defines a new universality
class while the E-L transition line belongs to the universality class
of the Harper equation. In addition, we show that
there exist special points in the interior as well as
on the boundary of the C phase which correspond to many new universality
classes. They are associated with higher order limit cycles of the RG
equations.
Our detailed analysis
suggests that the rest of the interior of the C phase is described by
a unique universal strange attractor of the RG flow.

In Section II, we describe the model and briefly review previously known
features of the phase diagram of the TBM describing Bloch electrons
with NNN interaction and make a comparision with the quantum spin problem.
The decimation scheme is reviewed in Section  III.
Section IV concentrates on higher order symmetric limit cycles, and
in Section V we describe the phenomenon of shifted
symmetry and doubling of the period of a limit cycle. Evidence for a
strange set is provided in Section VI. Finally, Section VII contains
a summary of our conclusions.

\section{ Tight Binding Model for Bloch Electrons in a Field }
The TBM describing Bloch electrons on a square lattice in a magnetic field
with both NN hoppings $t_a$ and $t_b$ and NNN hoppings $t_{ab}=
t_{a\bar{b}}$ is\cite{Thou}
\begin{eqnarray}
\{ t_a+2t_{ab} cos[2 \pi (\sigma(i+\frac{1}{2})+\phi)]\} \psi_{i+1}\nonumber\\
+ \{ t_a+2t_{ab}cos[2 \pi (\sigma(i-\frac{1}{2})+\phi)] \} \psi_{i-1}
\nonumber\\
+ 2t_b cos[2 \pi (\sigma i+\phi)] \psi_i
=E\psi_i
\end{eqnarray}
Here, $\sigma$ is the magnetic flux which we choose to be the
inverse golden
mean $\sigma=(\sqrt5-1)/2$. This TBM
was studied in detail in ref. \cite{Thou}.
Fig. 1 shows the phase diagram of the model in the space of the
parameters $\lambda=\frac{t_b}{t_a}$ and
$\alpha=2\frac{t_{ab}}{t_a}$. The Harper equation
corresponds to the limit $\alpha=0$ where the NNN coupling term is zero.
The phase diagram was obtained\cite{Thou} using
analytical methods to obtain the scaling behavior of the total
band width (TBW) and the Lyapunov exponents and carrying out a numerical
multifractal analysis.
The lines AC (E-C transition) and CE (C-L transition)
were found to be bicritical,  i.e. the
TBW scaled with the system size with the exponent $\delta=2$.
This was in contrast with the critical line BC separating the
E and L phases where
the exponent was known to be unity. Within the region bounded by the lines
AC and CE and the $\alpha$-axis, where the NNN coupling dominated, the
multifractal analysis did not lead to conclusive
results on the universality. This was due to lack in convergence of
the $f(\alpha)$ curve with the size of the system.
Furthermore, their numerical calculation of the TBW
was complicated by oscillatory terms superimposed on the power law.
However, this regime was conjectured to be critical.

The existence of the fat C phase implies
that the presence of the NNN coupling in the Bloch electron
problem introduces new universality classes. The same does not
happen if the cosine term in the Harper equation is just
replaced by more generic periodic functions.\cite{Eharper}

Eq. (1) involving both diagonal and off-diagonal disorder bears
some resemblance to the TBM describing quasiparticle fermion excitations
in a quantum XY spin chain.\cite{IIS}
Unlike Harper, both the anisotropic XY spin chain and Eq. (1) exhibit
C phase in a finite parameter interval. However, in the
TBM (1) the fat C phase is observed beyond a critical value of the NNN
hopping whereas in the spin model the fat C phase can be seen even
with infinitesimal spin space anisotropy.

Motivated by the success of our decimation scheme to confirm the existence
of a fat C phase and obtaining the universal behavior in the quantum
spin chain \cite{KS},
we now apply the method to Eq. (1). Our main focuss
is to obtain the universality classes of the fat C phase.

\section{ Decimation Scheme }

Our decimation approach describes the scaling properties of the wave
functions for a specific value of energy. Although all quantum states
are fractal
, it is more useful to study states
at the band edges.
This is because
the self-similar
behavior is usually observed only for the minimum and maximum energy states
and also for the band center if $E=0$
is an eigenenergy.
In order to get the overall picture,
it is sufficient to consider only one of these states and we will focuss on
the quantum state corresponding to $E_{min}$.

In RG analysis, in addition to fixing the quantum state
 , one has to also fix the phase factor $\phi$ in Eq. (1).
It has been pointed out in the previous
studies\cite{Ostlund},\cite{KS}, the wave function $\psi_i$
obtained by iterating the TBM diverges unless the phase factor $\phi$ is
tuned to some
critical value. In the Harper model, the critical value of the phase
factor is $\frac{1}{2}$ for the negative band edge.
For this value, the main peak is
centrally located and the wave function is symmetric about $i=0$.
In the study of the quantum spin model\cite{KS}, the phase factor had to be
varied
continuously in the fat C phase so that the main peak could be
centrally
located and the resulting wave function became bounded.
Determination of
this critical phase factor was essential in order to find the RG limit cycles
and to compute the universal scaling ratios. In general the phase factor
$\phi$ for obtaining symmetric wave function need not be identical to the
phase factor resulting in bounded wave functions.\cite{KS}

We consider an infinite lattice which extends in both positive
and negative directions from the $i=0$ site.
In the decimation scheme, all sites except those labelled by
 positive as well as
negative Fibonacci numbers are decimated. The resulting TBM
connecting the wave function
$\psi$ at two neighboring Fibonacci sites can be written as
\begin{eqnarray}
\psi(i+F_{n+1})&=&c^+_n(i)\psi(i+F_n)+d^+_n(i)\psi(i)\\
\psi(i-F_{n+1})&=&c^-_n(i)\psi(i-F_n)+d^-_n(i)\psi(i).
\end{eqnarray}
The index $n$ above refers to the level of decimation.

For the Harper
model it suffices to define only one set of the "decimation
functions" $c_n(i)$ and $d_n(i)$ because of the symmetry of the wave
function about $i=0$. This implies that
the scaling ratios are the same on the
positive and negative side. This was also asymptotically true in
the quantum spin model everywhere else except along the C-L
transition line\cite{KS}, where the wave functions were asymmetric about
$i=0$ resulting
in vanishing scaling ratios on one side.\cite{footnote}
The reason why we now have to introduce separate decimation functions
for the positive ($+$) and negative ($-$) side is that sometimes
the asymptotic ($n\to \infty$) $c_n^+(0), d_n^+(0)$ appear to be
shifted compared to $c_n^-(0), d_n^-(0)$.
This type of shifted symmetry turns out to be helpful in
locating higher order limit cycles and is discussed in detail in Section V.

Using the defining property of the Fibonacci numbers,
$F_{n+1} =F_n + F_{n-1}$, the following recursion relations
are obtained for $c_n$ and $d_n$ (we will omit the $+,-$ indices if
the equations do not depend upon them)\cite{Ketoja},\cite{KS}:
\begin{eqnarray}
c_{n+1} (i)&=& c_n(i+F_n) c_{n-1} (i+F_n)-d^{-1}_n(i) d_{n+1} (i)\\
d_{n+1} (i)&=& -d_n (i)
[d_n (i+F_n )+\nonumber \\
& & c_n(i+F_n) d_{n-1} (i+F_n)] c_n^{-1} (i).
\end{eqnarray}
For a fixed $i$, the above coupled equations for the decimation
functions define a RG flow which asymptotically ($n \rightarrow \infty$)
converge on an attractor.
In our earlier studies, the E, C, and L phases
were distinguished by the distinctions in the attractors of the RG flow.
In the Harper as well as in the quantum spin case, the C phase was
characterized by a $nontrivial$ asymptotic $p$-cycle
at the band edges with $p$ equal to $3$ or $6$.

The existence of a nontrivial $p$-cycle for the decimation functions often
implies
that the wave function is neither extended nor localized
and exhibits the self-similarity described by
\begin{equation}
\psi(i) \approx \psi([\sigma^pi+1/2])
\end{equation}
where $[\;\;]$ denotes the integer part. The $p$-cycle of the self-similar
bounded wave function can be used to define the universal scaling ratios
\begin{equation}
\zeta_j = \lim_{n \rightarrow \infty} |\psi(F_{pn+j})/\psi(0)|;\;\;j=0,...,p-1.
\end{equation}
This equation describes the decay of the wave function
with respect to the central peak.
A well-defined limit $\zeta_j$ exists for an integer $p$ for which
asymptotically $\psi(F_{n+p})\approx \psi(F_n)$. For an even $p$,
it often happens that $|\psi(F_{n+p/2})|\approx |\psi(F_n)|$ so that
the above equation defines actually only $p/2$ different scaling ratios.
Whenever any scaling ratio $\zeta_j$ takes a finite
value between zero and unity, the wave function over the
infinite system is neither localized nor extended. In order
to fully characterize the self-similarity of a wave function,
an infinite number of scaling ratios have to be defined. \cite{KS}
However, using the ones defined above one can already separate
different universality classes from each other.

The numerics to demonstrate a
$p$-cycle was rather challenging for the case where $E=0$ was not an
eigenenergy. This was because at many points in the phase diagram, the energy
was required with
16 digits precision (machine
double precision) in order to see the asymptotic cycle
with two or more digits of precision. Even for tridiagonal matrices, we
were able to determine the energy only up to $12$ digits. With this
precision, the decimation equations could be iterated only about $16$ times.
The conjecture for the existence of a limit cycle provided a very efficient
Newton method where the energy and the limit cycle were determined
self-consistently. Diagonalization
routines provided a good starting value of the energy which in principal could
be improved to an arbitrary precision. At many points where the transients
were rather long, the quadruple precision was used to confirm
the existence of a limit cycle.

\section{ Higher Order Symmetric Limit Cycles  }
Fig. 1 shows the E, C, and L phases of the model
in the $\lambda-\alpha$ space. The iteration of decimation equations
shows that the BC critical line defining the boundary between the E and L
phases is described by a $3$-cycle of the RG flow.
On this line (with the exception of the point C) the decimation functions
flow to the same limit cycle as for the critical Harper equation (point
B). However, the point C
is described by a different $3$-cycle defining a new universality class.
Table I compares the universal scaling ratios
in these two universality classes. Fig. 2(a-b)
show the wave function in these two cases. In both cases,
a bounded wave function was obtained for $\phi=1/2$ and the wave function
as well as decimation functions were symmetric about the center of the lattice:
i.e. $c^+=c^-$ and $d^+=d^-$.

The bicritical line AC was found to exhibit two new universality
classes: At the point $A$, the decimation functions were found to
asymptotically
converge to a period-$6$ limit cycle for $\phi=1/2$.
The regime bounded by the points A and C
on the line AC (excluding the points A and C) appears to be
described by a unique symmetric
limit cycle of the period $12$ for $\phi=1/2$. The $12$-cycle is
particularly
clear for the middle point M ($\lambda=.5, \alpha=1$).
However, for other points on this line, the RG flow may exhibit long
transients before settling on the limit cycle of the point M.
This is illustrated in Fig. 3 for the
point MM ($\lambda=.25$, $\alpha=1$). The RG iterations begin close
to the $6$-cycle of the point A and then follow the $3$-cycle
of the point C for a while before approaching
the $12$-cycle of the point M . There is additional
complexity involved because in order to observe
the approach we had to shift the data for MM by $6$ decimation levels.
We used the quadruple precision to confirm that the MM limit cycle
asymptotically converged to the limit cycle of M.

Symmetric period-$12$ cycles were also observed at the points F, G, and
H which fall on the line $t_a=t_{ab}$. However, the limit cycle at each point
was different implying different scaling properties and hence
different universality classes.

It should be noted that with the exception of the points B, C, and H,
all the limit cycles observed for $\phi=1/2$ resulted in asymptotically
 diverging
wave functions (see Fig. 2). The decimation functions corresponding to
diverging wave functions cannot be used to obtain universal scaling ratios
for the new universality classes. In section V, we determinine the
bounded wave functions by tuning the phase factor.

We also studied systematically other points on the lines CE and FH and
points inside the C phase, but we
did not see any evidence of limit cycles for $\phi=1/2$. Since the numerics
should be sufficient to show at least cycles of length $12$ (unless
there are very very long
transients), we can conjecture that there are no
cycles of the order 12 or shorter on these lines with the phase $1/2$.
In section VI, we present some evidence that the rest of the C phase
is described by a strange attractor of the RG flow.

\section{Shifted Symmetry and Period Doubling}

Fig. 4 shows the bounded wave functions obtained by tuning the phase
factor $\phi$ to a certain critical value.
Unlike the symmetric wave functions of section IV, these wave functions
are not symmetric about $i=0$. It turns out that this asymmetry is
due to a constant shift between the wave functions on the positive and
negative sides. Numerical iteration of the TBM (1) shows that in
addition to Eq. (7),
the wave functions on the positive and negative sides are asymptotically
related by
\begin{equation}
\psi(F_n)\approx \psi(-F_{n+s})
\end{equation}
Instead of the symmetric solution for the
decimation functions corresponding to $c^+_n (0)=c^-_n (0)$,
$d^+_n (0)=d^-_n (0)$ for all decimation levels $n$,
we found the asymptotic equations
\begin{eqnarray}
c^+_{n} (0)& \approx &c^-_{n+s} (0)\\
d^+_{n} (0)& \approx &d^-_{n+s} (0),
\end{eqnarray}
i.e. asymptotically there was a shift of $s$ levels between the
positive and negative decimation functions. Morover, in above
$+$ and $-$ could be interchanged with the same shift $s$, which
implied that
$c^+_{n} (0) \approx c^-_{n+s} (0) \approx c^+_{n+2s} (0)$,
i.e. the asymptotic period $p=2s$.
The shift $s$ was found to be equal to the period of the symmetric
limit cycle discussed in section IV.
Therefore, the phenomenon of shifted symmetry resulted in
doubling the period of a limit cycle for the decimation functions.

The points F and G, which
respectively fall on the intersection of the line $t_a=t_{ab}$ and
the lines $t_b=0$ and the self-dual line BD, exhibit the phenomenon
of shifted symmetry with $s=12$ (see Table II).
This value implies the asymptotic
cycle-length $24$.

The phenomenon of shifted symmetry is very crucial in locating the
limit cycles of period $24$. This is because with double precision
arithematics, the RG equations can be iterated only about $24$ times.
Without having the shifted symmetry we could not
have deduced that the asymptotic period is $24$ as
we could not go to big enough decimation levels to see
the full cycle on one side  only. This is shown explicitly
in Table II. The same shift and period is
observed for the critical phase at the point M in the middle of
the line AC.
At the point $A$ the shift is $6$ and the asymptotic period therefore
$12$.

\section{ Strange Set of the Renormalization Flow}

We explored the idea of describing the region bounded by the lines AC and
CE by a strange set of the RG flow. Fig. 5 shows
the two-dimensional projection of the attractor obtained by
plotting an inverse decimation function for two subsequent decimation
levels in various parts of the phase diagram.
Having $E_{min}$ up to $12$ digits, the RG equations
are estimated to give correct decimation functions up to
about $16$ levels. Transients were taken into account by excluding
the first six decimation levels from the data.
It is interesting to note that the iteration of the decimation functions
in three different parameter regimes namely
the line FH, the region CHG (excluding the line CH), and the region ACGF
appear to asymptotically converge on roughly the same set. Similar
figures were obtained also on the line CG, GD and the lines FG and GH. The
fact that different parts of the phase diagram are described by
similar invariant sets makes us to exclude the possibility that the observed
behavior is due to long transients. However, although the possibility of a very
long limit cycle cannot be completely ruled out, we believe that the observed
behavior
suggests that the interior of the fat C phase ( excluding the special
points which exhibit limit cycle ) is attracted to
a unique invariant set of the RG equations. We conjecture that the set
is a strange attractor.

The iterates of the RG flow on the CE line
seem to lie on the inner boundary of the invariant set
corresponding to the interior of the C phase (see Fig. 5). In the previous
studies
\cite{Thou}, the CE line defining the boundary of C and L phases
was found to be bicritical. In analogy with AC line, we would expect
that the decimation functions on the CE line converge to a limit cycle
of the order $24$ for $\phi=1/2$. We did not see
any evidence of this cycle. However, its existence can not be ruled out
specially in view of the possibility of long transients and the fact
that even the symmetric limit cycle could be of order $24$. Therefore,
the problem of determining the universality class along the CE line
describing the C-L transition remains open.

\section{Conclusions}
In this paper, we demonstrate that our decimation scheme is an extremely
useful tool to study general quasiperiodic TBMs.
The Bloch electron with NNN interaction
and the anisotropic quantum XY chain in a transverse field are two known
examples where the C phase exists in a finite parameter range.
In the spin problem, the C phase was characterized by four different
limit cycles of the RG flow.\cite{KS} The present study shows that in
the Bloch electron case the situation is lot more complex: in
addition to six
new universal limit cycles, which correspond to self-similar wave functions,
there is strong numerical evidence of a RG strange set. To best of our
knowledge,
this is the only example of a quasiperiodic model where the golden mean
incommensurability does not result in self-similar wave functions
at the band edges.  Even in the regime where no limit cycles exist,
the RG scheme provides a clear distinction between the E, C, and L
phases: in the E and L phases, the decimation functions are trivial, and in the
C
phase they assume finite non-trivial values.\cite{KS}
It should be noted that in this regime where the C phase is not described
by the limit cycles of the RG equations, the previous studies\cite{Thou}
showed lack of convergence in $f(\alpha)$ curve.

The general case of TBM (1) where the NNN interactions $t_{ab}$ and
$t_{a\bar b}$ are not equal provides an interesting limit
of the Bloch electron on a triangular lattice.\cite{tri} Our
studies have shown that\cite{CSK} the universality class of this model is
related to the subconformal universality class of the Ising model.
\cite{KS} This establishes an interesting relationship between
the anisotropic
Bloch electron with NNN interaction and anisotropic quantum spin
chains.

In the quantum spin problem, the fattening of the C phase is due
to the broken O(2) symmetry in spin space which translates to
the broken U(1) symmetry in the quasiparticle fermion Hamiltonian.
For the Bloch electron, the existence of C phase is due to
a NNN interaction. Although at this point we are unable to pin point
the commonality between the symmetric breaking in these two problems,
we believe that the origin of the fat C phase and new universality classes
 may be tied to certain
broken symmetries of the models.

\acknowledgements

The research of IIS is supported by a grant from National Science
Foundation DMR~093296. JAK is grateful for the hospitality
during his visit to the George Mason University.
IIS would like to acknowledge the hospitality of National Institue
of Standard and Technology where part of this work is done.

\begin{figure}
\caption{ The phase diagram of the anisotropic electron gas with $t_{ab}=
t_{a\overline{b}}$. The solid lines BC, AC, and CE are respectively the
E-L, E-C,
and C-L transition lines. With the exception of the point C, the BC line
is described by the Harper universality class.
 The bicritical line AC is described by three different
universal limit cycles corresponding to
the points A, C, and the regime in between A
and C. In addition, the points F, G, and H are also described by limit
cycles of the RG flow. The period of the limit cycle ( see section IV and
V ) is indicated in a
bracket close to the point. The two entries inside the bracket
describe the symmetric and the shifted-symmetry periods.
For example, (12) near the point H shows that it exhibits only the symmetric
limit cycle (with bounded wave function) with period $p=12$. The
(12,24) near the point G shows that it exhibits both the symmetric as
well as the
shifted-symmetry limit cycles of periods 12 and 24, respectively.}
\label{fig1}
\end{figure}

\begin{figure}
\caption{(a-g) shows the wave function at the points B, C, A, M, F, G,
and H
corresponding to the phase factor $\phi=1/2$ such that the reflection symmetry
about $i=0$ is preserved.
This causes
the wave function to diverge for the points A, M, F, and G while the wave
function at the point B, C, and H is bounded. In these
plots, the maximum value of the wave function is scaled to unity.}
\label{fig2}
\end{figure}

\begin{figure}
\caption{(a) The inverse decimation function $1/c_n (0)$ vs. $n$
($\phi=1/2$)
along the bicritical line AC: The data for the point MM ($\lambda=.25,
\alpha=1$; shown by crosses), shifted by
$6$ decimation levels,
eventually follows the $12$-cycle of the point $M$ ($\lambda=.5, \alpha
=1$; solid line with
small crosses showing the locations of the periodic orbit).}
\label{fig3}
\end{figure}

\begin{figure}
\caption{(a-d) show the absolute value of the wave function
at the points A ($\phi=1/4$), M ($\phi=.3202185$) , F ($\phi=.2777...$),
and G ($\phi= 1/3$) . The phase factor $\phi$ is chosen
so that the main peak is centrally located resulting in a bounded wave
function.}
\label{fig4}
\end{figure}

\begin{figure}
\caption{ A two-dimensional projections of the inverse decimation functions
inside the fat C phase.
The data has been obtained by sampling on the line FH (a), in the region CHG
(b),
and
(c) of the whole C phase which includes (a) and (b) and also the square ACGF
and the points above it. The dark
dots correspond to the data obtained along to the line CE
(excluding the points C and H).
}
\label{fig5}
\end{figure}

\begin{table}
\caption{ The
universal scaling ratios $\zeta_j$ $(j=1,3)$
at the point $B$ (Harper) and at the point $C$
(bicritical). }
\begin{tabular}{cccccc}
 & & & & &\\
$j$ &  $\zeta_j (B)$ & & $\zeta_j (C)$   \\
 & & & & &\\
\tableline
 & & & & &\\
0 (0, 2, 8, 34, 144, 610,...) & 0.2107 & & 0.1712 \\
1 (1, 3, 13, 55, 233,... ) & 0.2107 & &  0.2353\\
2 (1, 5, 21, 89, 377,... ) & 0.2107 & & 0.2387\\
 & & & & &\\
\end{tabular}
\label{table1}
\end{table}

\begin{table}
\caption{ The decimation functions for the point G ($\phi=1/3$)
at site $i=0$ showing the shifted symmetry with $s=12$
and thus indirectly implying the limit cycle of length $24$.
We see that $c^+_n(0) \approx c^-_{n\pm 12}(0)$ and also
$d^+_n(0) \approx d^-_{n\pm 12}(0)$.}

\begin{tabular}{cccccc}
 & & & & &\\
$n$ &  $c^+_n(0)$  & $c^-_n(0)$  &$d^+_n(0)$ & $d^-_n(0)$\\
 & & & & &\\
\tableline
 & & & & &\\
6 & -0.817  &  -2.477 & -3.499E-02 &    0.109\\
7 &  2.852  & -6.304  & -1.257     &   -3.952   \\
8 & -4.149  &  5.509  &  1.439    &    -0.501   \\
9 &  1.990  &  1.176  &  0.191   &  -4.571E-02\\
10 &  -4.463 &  -1.166 &   -1.685  &   0.302 \\
11 &   2.733 &  -4.122 &  -1.168  &   -1.808  \\
12 &  -3.255 &   42.235 &   0.498  &    -9.109  \\
13 &   1.517 &   1.559 &   0.140  &  -1.328E-02\\
14 &   4.982 &   -6.744&   3.607   &  0.250   \\
15 &  -3.738 &   -0.831&  -0.427  &  -1.720E-02\\
16 &  -1.961 &   -1.957&   0.467  &  8.547E-02  \\
17 &  -12.536 &    9.281&   -6.087  &   0.944  \\
18 &  -2.507 &   -0.809&   0.109  &  -3.454E-2\\
19 &  -6.266 &    2.874&  -3.942   &  -1.26   \\
20 &   5.532 &   -4.132& -0.502    &   1.436  \\
21 &   1.174 &    1.996&  -4.567E-02 &  0.191\\
22 &  -1.168 &   -4.457 &   0.303   & -1.684  \\
23 &  -4.119 &     2.738&  -1.807    & -1.168\\
24 &  42.288 &    -3.256 & -9.106   & 0.498 \\
& & & & & \\
\end{tabular}
\label{table2}
\end{table}


\begin{references}
\bibitem{footnote}The C-L transition line in the
quantum spin model is special as it also corresponds to
the onset of magnetic long range order and to the conformal
invariance of the system.\cite{IIS}
\end{references}
\end{document}